\documentclass[12pt]{article}
\usepackage{amsfonts}
\usepackage{amsmath,amssymb}
\usepackage{bbm}
\newcommand{\CA}{\mathcal{A}}    
\newcommand{\CG}{\mathcal{G}}    
\newcommand{\CL}{\mathcal{L}}    
\newcommand{\CN}{\mathcal{N}}    
\newcommand{\CO}{\mathcal{O}}    
\newcommand{\CP}{\mathcal{P}}    

\newcommand{\CT}{\mathcal{T}}    
\newcommand{\CE}{\mathcal{E}}    
\newcommand{\frS}{\mathfrak{S}}    
\newcommand{\FR}{\mathbbm{R}}     
\newcommand{\FC}{\mathbbm{C}}     
\newcommand{\CPP}{{\mathbbm{C}P}}    
\newcommand{\dpar}{\partial}     
\newcommand{\dparb}{{\bar{\partial}}}     
\newcommand{\diag}{{\mathrm{diag}}}     
\newcommand{\ed}{{\dot{1}}}    
\newcommand{\zd}{{\dot{2}}}    
\newcommand{\ald}{{\dot{\alpha}}}     
\newcommand{\bed}{{\dot{\beta}}}     
\newcommand{\gad}{{\dot{\gamma}}}     
\newcommand{\eps}{{\varepsilon}}     
\newcommand{\eand}{{~~~\mbox{and}~~~}}     
\newcommand{\der}[1]{\frac{\dpar}{\dpar #1}}   
\newcommand{\tr}{\,\mathrm{tr}\,}     

\newcommand{\agl}{\mathrm{gl}}     
\newcommand{\sGL}{\mathrm{GL}}     

\newcommand{\remark}[1]{}     

\begin{document}
\begin{center}
\Large The Mini-Superambitwistor Space\footnote{Talk given at the
International Workshop ``Supersymmetries and Quantum Symmetries'' (SQS'05),
Bogoliubov Laboratory of Theoretical Physics, JINR, Dubna, 27-31 July 2005.} \\[0.2cm]
\large Christian S\"{a}mann\\[0.2cm]
\normalsize
{\em Institut f\"{u}r Theoretische Physik}\\
{\em Universit\"{a}t Hannover}\\
{\em Appelstra{\ss}e 2, D-30167 Hannover, Germany}\\[0.2cm]
{\tt saemann@itp.uni-hannover.de}
\end{center}

\begin{quote}
We present the construction of the mini-superambitwistor space,
which is suited for establishing a Penrose-Ward transform between
certain bundles over this space and solutions to the $\CN=6$ super
Yang-Mills equations in three dimensions.
\end{quote}

The essential point underlying twistor string theory
\cite{Witten:2003nn} is the marriage of Calabi-Yau and twistor
geometry in the space $\CPP^{3|4}$. This complex projective space
is a Calabi-Yau supermanifold and simultaneously the supertwistor
space of the complexified, compactified Minkowski space. The
interest in a twistorial string theory is related to the fact that
twistor geometry allows for a very convenient description of the
solution spaces of classical gauge theories
\cite{Penrose:ca,Ward:vs}. In such a description, spacetime is
associated with a certain complex manifold, its {\em twistor
space}. Subsequently, the space of holomorphic vector bundles over
this twistor space is mapped to the solution space of the gauge
theory via a so-called {\em Penrose-Ward transform}.

Suitable twistor spaces are well-known for four-dimensional
self-dual Yang-Mills theory and its supersymmetric extensions, as
well as for four-dimensional Yang-Mills theory and its $\CN=3$
supersymmetric extension, see
\cite{Witten:1978xx,Isenberg:kk,Witten:2003nn,Popov:2004rb}. Via
dimensional reduction, one obtains so-called {\em mini-twistor
spaces} upon which a description of the solution space to the
Bogomolny monopole equation \cite{Hitchin:1982gh} and its
supersymmetric extensions can be constructed, see e.g.\
\cite{Chiou:2005jn,Popov:2005uv}. In this collection, the
mini-twistor space suited for a Penrose-Ward transform yielding
solutions to the three-dimensional Yang-Mills-Higgs equations and
their $\CN=6$ supersymmetric extension is evidently missing. This
gap was filled in \cite{Saemann:2005ji}, and here we will
concisely report on the results and thus review the construction
of the {\em mini-superambitwistor space}.

Let us first recall that twistors were invented by Penrose to give
a unified description of General Relativity and quantum mechanics.
Consider a light ray, which is given by the set of points
$x^{\alpha\ald}$ satisfying the equation
$x^{\alpha\ald}=x_0^{\alpha\ald}+tp^{\alpha\ald}$. Here,
$x_0^{\alpha\ald}$ is an arbitrary point on the light ray and
$t\in \FR$ is a parameter. Taking a light ray which does not pass
through the origin, one can obviously choose $x_0^{\alpha\ald}$ to
be null. Since one can decompose every null vector into a pair of
commuting two-spinors, we can rewrite the equation defining our
light ray as
$x^{\alpha\ald}=c\omega^\alpha\tilde{\omega}^\ald+t\lambda^\ald\tilde{\lambda}^\alpha$.
Multiplication of this equation by $\lambda_\ald$ together with
the right choice of normalization
$c=(\tilde{\omega}^\ald\lambda_\ald)^{-1}$ gives rise to the {\em
incidence relation}
\begin{equation}\label{incidence1}
\omega^\alpha\ =\ x^{\alpha\ald}\lambda_\ald~.
\end{equation}
A {\em twistor} $Z^i$ is now a projective\footnote{the incidence
relation is invariant under scaling} pair of two-spinors
$Z^i=(\omega^\alpha,\lambda_\ald)\in \CPP^3$, which transforms
under coordinate shifts $x^{\alpha\ald}\rightarrow
x^{\alpha\ald}+r^{\alpha\ald}$ according to the incidence
relation.

The space $\CPP^3$ is the twistor space of the complexified,
compactified Minkowski space $S^4_c$. Taking out the sphere
$S^2\cong \CPP^1=\{(\omega^\alpha\neq 0,\lambda_\ald=0)\}$ which
is incident to $x^{\alpha\ald}=\infty$, we can consider
$\lambda_\ald$ as the homogeneous coordinates on the Riemann
sphere $\CPP^1$. Due to the incidence relation, the
$\omega^\alpha$s are homogeneous polynomials of degree one in
$\lambda_\ald$ and thus describe a section of the rank two complex
vector bundle $\CO(1)\oplus\CO(1)\rightarrow \CPP^1$. This
bundle's total space, which we denote by $\CP^3$, is the twistor
space of $\FC^4$.

The incidence relation allows us furthermore to establish the
double fibration
\begin{equation}\label{dblord}
\begin{aligned}
\begin{picture}(80,40)
\put(0.0,0.0){\makebox(0,0)[c]{$\CP^{3}$}}
\put(64.0,0.0){\makebox(0,0)[c]{$\FC^{4}$}}
\put(32.0,33.0){\makebox(0,0)[c]{$\FC^4\times \CPP^1$}}
\put(7.0,18.0){\makebox(0,0)[c]{$\pi_2$}}
\put(55.0,18.0){\makebox(0,0)[c]{$\pi_1$}}
\put(25.0,25.0){\vector(-1,-1){18}}
\put(37.0,25.0){\vector(1,-1){18}}
\end{picture}
\end{aligned}
\end{equation}
where
$\pi_2(x^{\alpha\ald},\lambda_\ald)=(x^{\alpha\ald}\lambda_\ald,\lambda_\ald)$
and $\pi_1(x^{\alpha\ald},\lambda_\ald)=(x^{\alpha\ald})$, from
which one can easily read off the following twistor
correspondence: A point $x^{\alpha\ald}\in\FC^4$ defines a sphere
$S^2\cong\CPP^1_x$ embedded in $\CP^3$, while a point $p\in\CP^3$
is incident to a null two-plane $\FC^2_p$ in $\FC^4$.

{}To obtain an $\CN$-extended supertwistor space, one can simply
start from the projective superspace $\CPP^{3|\CN}$, take out the
$\CPP^{1|\CN}$ corresponding to the point at infinity and arrive
at the supervector bundle
\begin{equation}
\CP^{3|\CN}\ :=\ \FC^2\otimes \CO(1)\oplus \FC^\CN\otimes \Pi\CO(1)
\end{equation}
over the Riemann sphere $\CPP^1$. The operator $\Pi$ simply
inverts the parity of the fibre coordinates of a vector bundle,
and one has therefore $\CN$ additional homogeneous Gra{\ss}mann
coordinates $\eta_1,\ldots,\eta_\CN$. The incidence relation
\eqref{incidence1} is extended to
\begin{equation}\label{incidence2}
\omega^\alpha\ =\ x^{\alpha\ald}\lambda_\ald\eand
\eta_i\ =\ \eta_i^\ald\lambda_\ald~,
\end{equation}
which naturally gives rise to the double fibration
\begin{equation}\label{dblsuper}
\begin{aligned}
\begin{picture}(80,40)
\put(0.0,0.0){\makebox(0,0)[c]{$\CP^{3|\CN}$}}
\put(64.0,0.0){\makebox(0,0)[c]{$\FC^{4|2\CN}$}}
\put(32.0,33.0){\makebox(0,0)[c]{$\FC^{4|2\CN}\times \CPP^1$}}
\put(7.0,18.0){\makebox(0,0)[c]{$\pi_2$}}
\put(55.0,18.0){\makebox(0,0)[c]{$\pi_1$}}
\put(25.0,25.0){\vector(-1,-1){18}}
\put(37.0,25.0){\vector(1,-1){18}}
\end{picture}
\end{aligned}
\end{equation}
Here, $\pi_2$ is given by the extended incidence relations
\eqref{incidence2} and $\pi_1$ is the obvious projection.

In the special case $\CN=4$, the first Chern number of (the
tangent bundle of) the supertwistor space vanishes. This is due to
the fact that Berezin integration is equivalent to a
differentiation and therefore the contribution of $\Pi\CO(1)$ to
the total first Chern number is $-1$. Altogether, we have a
contribution of $2$ from the tangent bundle to the Riemann sphere
$TS^2\cong \CO(2)$ and $1$ from each bosonic $\CO(1)$, which is
cancelled by the $-4$ of the fermionic line bundles $\Pi\CO(1)$.
Thus, $\CP^{3|4}$ comes with a holomorphic measure
$\Omega^{3,0|4,0}$ and this supertwistor space is a Calabi-Yau
supermanifold.

One can now establish a relation between a topological string
theory having $\CP^{3|4}$ as its target space and $\CN=4$
self-dual Yang-Mills theory in four dimensions: The open
topological B-model on $\CP^{3|4}$ with a stack of $n$
space-filling D5-branes is equivalent to holomorphic Chern-Simons
theory on the same space, which describes the dynamics of a
$\agl(n,\FC)$-valued connection $(0,1)$-form $\CA^{0,1}$ on a rank
$n$ complex vector bundle $\CE\rightarrow\CP^{3|4}$
\cite{Witten:2003nn}. The action of this holomorphic Chern-Simons
theory reads as \cite{Witten:2003nn}
\begin{equation}
S\ =\ \int \Omega^{3,0|4,0}\wedge \tr\left(
\CA^{0,1}\wedge\dparb\CA^{0,1}+\tfrac{2}{3}\CA^{0,1}\wedge\CA^{0,1}\wedge\CA^{0,1}\right)~,
\end{equation}
where $\Omega^{3,0|4,0}$ is the holomorphic measure on
$\CP^{3|4}$. (Some minor assumptions about the explicit form of
$\CA^{0,1}$ have to be made at this point, see
\cite{Popov:2004rb}.) The corresponding equations of motion are
given by $\dparb \CA^{0,1}+\CA^{0,1}\wedge\CA^{0,1}=0$ and their
solutions describe holomorphic structures which promote the
complex vector bundle $\CE$ to holomorphic vector bundles
$(\CE,\CA^{0,1})$. Via a generalized Penrose-Ward transform using
supertwistor spaces \cite{Witten:2003nn,Popov:2004rb}, one can map
these holomorphic vector bundles to solutions to the $\CN=4$
extended SDYM equations on $\FC^{4}$. These equations are the
supersymmetric extensions of the self-dual Yang-Mills equations
$F_{\mu\nu}=\frac{1}{2}\eps_{\mu\nu\rho\sigma}F^{\rho\sigma}$,
which read in spinorial notation $F_{\mu\nu}\rightarrow
F_{\alpha\ald\beta\bed}=\eps_{\alpha\beta}f_{\ald\bed}+\eps_{\ald\bed}f_{\alpha\beta}$
as
\begin{equation}\label{SDYMeom}
\begin{aligned}
f_{\ald\bed}&\ =\ 0~,\\
\nabla_{\alpha\ald}\chi^{\alpha i}&\ =\ 0~,\\
\Box\phi^{[ij]}&\ =\ +\tfrac{1}{2}\{\chi^{\alpha i},\chi^j_{\alpha}\}~,\\
\eps^{\ald\dot{\gamma}}\nabla_{\alpha\ald}\tilde{\chi}^{[ijk]}_\gad&\
=\ -2
[\phi^{[ij},\chi^{k]}_{\alpha}]~,\\
\eps^{\ald\dot{\gamma}}\nabla_{\alpha\ald}G^{[ijkl]}_{\dot{\gamma}
\dot{\delta}}&\ =\
-\{\chi^{[i}_{\alpha},\tilde{\chi}^{jkl]}_{\dot{\delta}}\}
+[\phi^{[ij},\nabla_{\alpha\dot{\delta}}\phi^{kl]}]~,
\end{aligned}
\end{equation}
where the nontrivial fields $(f_{\alpha\beta},\chi^i_\alpha,
\phi^{[ij]},\tilde{\chi}^{[ijk]}_{\ald},G^{[ijkl]}_{\ald\bed})$
have helicities $(+1,+\frac{1}{2},0,-\frac{1}{2},-1)$. Neglecting
the trivial field $f_{\ald\bed}$, the field content of $\CN=4$
self-dual Yang-Mills theory is identical to the one of $\CN=4$
super Yang-Mills theory, but the interactions in the two theories
are different.

Let us now turn our attention to another twistor space, the
so-called {\em superambitwistor space}, upon which a Penrose-Ward
transform for the full $\CN=3$ super Yang-Mills equations can be
constructed. Important here is the observation that the $\CN=3$
supermultiplet is reducible and splits into a self-dual and an
anti-self-dual part. One is thus naturally led to glue together
the twistor space $\CP^{3|3}$ for $\CN=3$ self-dual Yang-Mills
theory with a dual copy\footnote{The word dual here refers to the
spinor indices and {\em not} to the line bundles underlying
$\CP^{3|3}$.} $\CP^{3|3}_*$ for the anti-self-dual part. Denoting
the homogeneous coordinates on these two spaces by
$(\omega^\alpha,\lambda_\ald,\eta_i)$ and
$(\omega^\ald_*,\lambda_\alpha^*,\eta^i_*)$, we can write the
appropriate gluing condition as the quadric equation
\begin{equation}\label{quadric}
\kappa\ :=\ \omega^\alpha\lambda^*_\alpha-\omega^\ald_*\lambda_\ald+2\eta^i_*\eta_i\ =\ 0~,
\end{equation}
which defines the superambitwistor space $\CL^{5|6}$ as a subset
of $\CP^{3|3}\times \CP^{3|3}_*$.

{}To examine the geometry of $\CL^{5|6}$ more closely, note that
the appropriate incidence relations for the space $\CP^{3|3}\times
\CP^{3|3}_*$ read as
\begin{equation}
\omega^\alpha\ =\ x_R^{\alpha\ald}\lambda_\ald~,~~~
    \eta_i\ =\ \eta_i^\ald\lambda_\ald~,~~~
    \omega_*^\ald\ =\ x_L^{\alpha\ald}\lambda^*_\alpha~,~~~
    \eta^i_*\ =\ \eta_*^{i\alpha}\lambda^*_\alpha~.
\end{equation}
The quadric condition \eqref{quadric} is automatically and most
generally solved, if we choose
\begin{equation}
x_R^{\alpha\ald}\ =\ x^{\alpha\ald}-\eta_*^{\alpha
    i}\eta_i^\ald\eand x_L^{\alpha\ald}\ =\ x^{\alpha\ald}+\eta_*^{\alpha
    i}\eta_i^\ald~,
\end{equation}
and thus $x_R^{\alpha\ald}$ and $x_L^{\alpha\ald}$ are indeed
right- and left-handed chiral coordinates on the chiral
superspaces $\FC^{4|6}_R$ and $\FC^{4|6}_L$. These incidence
relations, too, define a double fibration:
\begin{equation}\label{ambidouble}
\begin{aligned}
\begin{picture}(85,40)
\put(0.0,0.0){\makebox(0,0)[c]{$\CL^{5|6}$}}
\put(64.0,0.0){\makebox(0,0)[c]{$\FC^{4|12}$}}
\put(32.0,33.0){\makebox(0,0)[c]{$\FC^{4|12}\times \CPP^1\times
\CPP^1_*$}} \put(7.0,18.0){\makebox(0,0)[c]{$\pi_2$}}
\put(55.0,18.0){\makebox(0,0)[c]{$\pi_1$}}
\put(25.0,25.0){\vector(-1,-1){18}}
\put(37.0,25.0){\vector(1,-1){18}}
\end{picture}
\end{aligned}
\end{equation}
Over the superambitwistor space, one can then establish a
Penrose-Ward transform which is a map between solutions to the
$\CN=3$ super Yang-Mills equations and certain holomorphic vector
bundles over $\CL^{5|6}$, see e.g.\ \cite{Popov:2004rb}.

One can also find twistor spaces which describe self-dual
Yang-Mills theory after a dimensional reduction $\FC^4\rightarrow
\FC^3$. In our conventions, one can make the following
identification of vector fields\footnote{See also
\cite{Popov:2005uv}, where the explicit identification is slightly
different.}
\begin{equation}
\CT_3\ :=\ \der{x^3}\ =\ -\der{x^{1\zd}}+\der{x^{2\ed}}\sim
\der{x^{[2\ed]}}~.
\end{equation}
Dimensional reduction of the $x^3$-direction thus implies
eliminating the modulus $x^{[\alpha\ald]}$. On the twistor space
side, this can be done by changing the incidence relation on
$\CP^3=\CO(1)\oplus\CO(1)$ from
$\omega^\alpha=x^{\alpha\ald}\lambda_\ald$ to
$\upsilon=x^{\ald\bed}\lambda_\ald\lambda_\bed$. The latter
equation defines sections of the line bundle $\CP^2:=\CO(2)$ over
the Riemann sphere $\CPP^1$. More formally, one has
$(\CO(1)\oplus\CO(1))/\CG=\CO(2)$, where $\CG$ is the abelian
group generated by the holomorphic vector field on $\CP^3$ which
corresponds to $\CT_3$ \cite{Popov:2005uv}. The space $\CP^2$ is
called the {\em mini-twistor space} \cite{Hitchin:1982gh}, and the
corresponding double fibration reads as
\begin{equation}\label{dblmini}
\begin{aligned}
\begin{picture}(80,40)
\put(0.0,0.0){\makebox(0,0)[c]{$\CP^{2}$}}
\put(64.0,0.0){\makebox(0,0)[c]{$\FC^{3}$}}
\put(32.0,33.0){\makebox(0,0)[c]{$\FC^3\times \CPP^1$}}
\put(7.0,18.0){\makebox(0,0)[c]{$\nu_2$}}
\put(55.0,18.0){\makebox(0,0)[c]{$\nu_1$}}
\put(25.0,25.0){\vector(-1,-1){18}}
\put(37.0,25.0){\vector(1,-1){18}}
\end{picture}
\end{aligned}
\end{equation}

After applying this reduction to the space $\CP^{3|3}\times
\CP^{3|3}_*$, we arrive at the space $\CP^{2|3}\times \CP^{2|3}_*$
together with the incidence relations
\begin{equation}\label{incidence3}
\upsilon\ =\ y^{\ald\bed}\lambda_\ald\lambda_\bed~,~~~
    \eta_i\ =\ \eta_i^\ald\lambda_\ald~,~~~
    \upsilon_*\ =\ y^{\ald\bed}_*\lambda^*_\ald\lambda^*_\bed~,~~~
    \eta^i_*\ =\ \eta_*^{i\ald}\lambda^*_\ald~.
\end{equation}
Here, we adjusted the spinor indices anticipating that there is no
distinction between left- and right-handed spinors on $\FC^3$. The
quadric equation \eqref{quadric} is correspondingly reduced to the
equation
\begin{equation}\label{quadred}
\left.\left(\upsilon-\upsilon_*+
         2\eta^i_*\eta_i\right)\right|_{\lambda=\lambda_*}\ =\
         0~,
\end{equation}
and this condition defines the {\em mini-superambitwistor space}
as a subset of $\CP^{2|3}\times \CP^{2|3}$ \cite{Saemann:2005ji}.
Altogether, we have the dimensional reductions
\begin{equation}
\begin{tabular}{ccc}
    $\CP^{3|3}\times \CP^{3|3}_*$ & $\ \rightarrow\ $ &
    $\CP^{2|3}\times \CP^{2|3}_*$\\
    $\downarrow$ &  & $\downarrow$\\
    $\CL^{5|6}$ & $\ \rightarrow\ $ & $\CL^{4|6}$\\
    $\downarrow$ &  & $\downarrow$\\
    $\CPP^1\times \CPP^1_*$ & & $\CPP^1\times
    \CPP^1_*$
\end{tabular}
\end{equation}
The reduced quadric equation \eqref{quadred} is solved, if we
choose the ``chiral coordinates''
\begin{equation}\label{ident}
y^{\ald\bed}\ =\ y_0^{\ald\bed}-\eta_*^{i(\ald}\eta_i^{\bed)}\eand
y_*^{\ald\bed}\ =\ y_0^{\ald\bed}+\eta_*^{i(\ald}\eta_i^{\bed)}
\end{equation}
in the incidence relations.

The incidence relations \eqref{incidence3} determine together with
the reduced quadric equation \eqref{quadred} yielding
\eqref{ident} the dimensional reduction of the double fibration
\eqref{ambidouble} to be
\begin{equation}\label{dbltwoinone}
\begin{aligned}
\begin{picture}(200,95)
\put(0.0,0.0){\makebox(0,0)[c]{$\CL^{4|6}$}}
\put(0.0,52.0){\makebox(0,0)[c]{$\CL^{5|6}$}}
\put(186.0,0.0){\makebox(0,0)[c]{$\FC^{3|12}$}}
\put(186.0,52.0){\makebox(0,0)[c]{$\FC^{4|12}$}}
\put(91.0,33.0){\makebox(0,0)[c]{$\FC^{3|12}\times\CPP^1\times\CPP^1_*$}}
\put(91.0,85.0){\makebox(0,0)[c]{$\FC^{4|12}\times\CPP^1\times\CPP^1_*$}}
\put(37.5,25.0){\vector(-3,-2){25}}
\put(145.5,25.0){\vector(3,-2){25}}
\put(37.5,77.0){\vector(-3,-2){25}}
\put(145.5,77.0){\vector(3,-2){25}}
\put(0.0,45.0){\vector(0,-1){37}}
\put(180.0,45.0){\vector(0,-1){37}}
\put(85.0,78.0){\vector(0,-1){37}}
\put(24.0,78.0){\makebox(0,0)[c]{$\pi_2$}}
\put(162.0,78.0){\makebox(0,0)[c]{$\pi_1$}}
\put(24.0,26.0){\makebox(0,0)[c]{$\nu_2$}}
\put(162.0,26.0){\makebox(0,0)[c]{$\nu_1$}}
\end{picture}
\end{aligned}
\end{equation}

Although all the constructions seem to go through without
difficulties, the geometry of $\CL^{4|6}$ contains some surprises.
First of all, note that the reduced quadric condition
\eqref{quadred} is not imposed over the whole base $\CPP^1\times
\CPP^1_*$ of the supervector bundle $\CP^{3|3}\times \CP^{3|3}_*$,
but only over its diagonal $\Delta:=\diag(\CPP^1\times \CPP^1_*)$,
which is the subspace of the base for which $\lambda=\lambda_*$.
Considering the projection $\pi:\CL^{4|6}\rightarrow \CPP^1\times
\CPP^1_*$, we see that $\pi^{-1}(\lambda,\lambda_*)\cong \FC^2$
for $\lambda\neq \lambda_*$, but $\pi^{-1}(\lambda,\lambda_*)\cong
\FC$ on the diagonal $\Delta$. One can in fact show that
$\CL^{4|6}$ is a fibration \cite{Saemann:2005ji}, but since its
fibre dimension varies, it is evidently not a vector bundle.
However, we will see in the following, that this seemingly
unpleasant property does not impose any relevant obstructions.

First, let us find an interpretation of the geometries involved in
the double fibration for the mini-superambitwistor space which is
contained in \eqref{dbltwoinone}. Recall that for the well-known
double fibrations \eqref{dblord}, \eqref{dblsuper} and
\eqref{ambidouble}, there is a nice interpretation in terms of
flag manifolds \cite{Ward:vs}. In the case of the double
fibrations for the mini-supertwistor and mini-superambitwistor
spaces, we find a quite similar description. For simplicity, we
restrict our discussion to the bosonic subspaces, i.e.\ to the
bodies of the considered superspaces.

After imposing reality conditions \cite{Popov:2005uv} on the
spaces involved in the double fibration \eqref{dblmini}, we obtain
\begin{equation}\label{minidouble}
\begin{aligned}
\begin{picture}(80,40)
\put(0.0,0.0){\makebox(0,0)[c]{$\CP^2_r$}}
\put(64.0,0.0){\makebox(0,0)[c]{$\FR^{3}$}}
\put(32.0,33.0){\makebox(0,0)[c]{$\FR^3\times S^2$}}
\put(7.0,18.0){\makebox(0,0)[c]{$\nu_2$}}
\put(55.0,18.0){\makebox(0,0)[c]{$\nu_1$}}
\put(25.0,25.0){\vector(-1,-1){18}}
\put(37.0,25.0){\vector(1,-1){18}}
\end{picture}
\end{aligned}
\end{equation}
The space $\FR^3\times S^2$ on the top is the space of oriented
lines in $\FR^3$ with one marked point. Keeping the point and
dropping the line evidently leads to an element of the space
$\FR^3$, while keeping the line and dropping the point -- or,
alternatively, moving the point as close as possible to the origin
-- leads to an element of $\CP^2_r=\CO(2)\cong TS^2$. The
projections $\nu_1$ and $\nu_2$ in \eqref{minidouble} have
therefore a clear geometric meaning.

The real double fibration for the bosonic part of the
mini-superambitwistor space,
\begin{equation}\label{miniambidouble}
\begin{aligned}
\begin{picture}(80,40)
\put(0.0,0.0){\makebox(0,0)[c]{$\CL^4_r$}}
\put(64.0,0.0){\makebox(0,0)[c]{$\FR^{3}$}}
\put(32.0,33.0){\makebox(0,0)[c]{$\FR^3\times S^2\times S^2_*$}}
\put(7.0,18.0){\makebox(0,0)[c]{$\nu_2$}}
\put(55.0,18.0){\makebox(0,0)[c]{$\nu_1$}}
\put(25.0,25.0){\vector(-1,-1){18}}
\put(37.0,25.0){\vector(1,-1){18}}
\end{picture}
\end{aligned}
\end{equation}
has a similar interpretation. The space $\FR^3\times S^2\times
S^2_*$ is the space of two oriented lines in $\FR^3$ with a common
marked point. Dropping the lines leads again to elements of
$\FR^3$, while moving the point on one of the lines (together with
the attached second line) to its shortest distance to the origin
yields an element of $\CL^4_r$.

Ultimately, one is certainly interested in extending the
discussion to the level of topological strings. Recall that the
superambitwistor space is in fact a (local) Calabi-Yau
supermanifold \cite{Witten:2003nn}, and one can therefore use
$\CL^{5|6}$ as a target space for the topological B-model.
Although the mini-superambitwistor space $\CL^{4|6}$ is not a
manifold, one nevertheless finds that a certain Calabi-Yau
property still persists.

A Calabi-Yau manifold can be defined as a manifold whose tangent
bundle has vanishing first Chern class. Chern classes of vector
bundles in turn are related to certain degeneracy loci of a set of
generic sections: On a rank $n$ vector bundle, the first Chern
class is Poincar{\'e} dual to the degeneracy loci of $n$ generic
sections. Straightforward calculations show, that the appropriate
degeneracy loci for $\CL^{5|6}$ and $\CL^{4|6}$ are rationally
equivalent \cite{Saemann:2005ji}. This is a strong hint that
$\CL^{4|6}$ comes with the necessary properties for using this
space as target space for a topological B-model. Furthermore, if
the conjecture \cite{Neitzke:2004pf} by Aganagic, Neitzke and Vafa
is correct and $\CL^{5|6}$ is indeed the mirror symmetry partner
of $\CP^{3|4}$ then by applying dimensional reduction, it is only
natural to conjecture that the mini-superambitwistor space
$\CL^{4|6}$ is the mirror of the mini-supertwistor space
$\CP^{2|4}$.

As far as the Penrose-Ward transform is concerned, the discussion
over $\CL^{4|6}$ follows essentially the lines of the discussion
over $\CL^{5|6}$. Since the mini-superambitwistor space is only a
fibration and not a manifold, we have to slightly extend the
notion of a vector bundle. We define an {\em $\CL^{4|6}$-bundle of
rank $n$} by a \v{C}ech 1-cocycle $\{f_{ab}\}\in
\check{Z}^1(\CL^{4|6},\frS)$, where $\frS$ is the sheaf of smooth
$\sGL(n,\FC)$-valued functions on $\CL^{4|6}$. This 1-cocycle
takes over the r{\^o}le of a transition function in an ordinary vector
bundle. A {\em holomorphic $\CL^{4|6}$-bundle} is correspondingly
defined by a holomorphic such \v{C}ech 1-cocycle. We call two
$\CL^{4|6}$-bundles given by two 1-cocylces $\{f_{ab}\}$ and
$\{f'_{ab}\}$ {\em topologically equivalent}, if there is a
\v{C}ech 0-cochain $\{\psi_a\}\in\check{C}^0(\CL^{4|6},\frS)$ such
that $f_{ab}=\psi_a^{-1}f'_{ab}\psi_b$. In particular, an
$\CL^{4|6}$-bundle is topologically trivial (topologically
equivalent to the trivial bundle), if its defining 1-cocycle can
be decomposed by a \v{C}ech 0-cochain according to
$f_{ab}=\psi^{-1}_a\psi_b$.

With these definitions, we can state that topologically trivial,
holomorphic $\CL^{4|6}$-bundles, which become holomorphically
trivial vector bundles upon restriction to any $\CPP^1\times
\CPP^1_*\subset\CL^{4|6}$ are equivalent to solutions of the
$\CN=6$ super Yang-Mills equations on $\FC^{3}$. The number of
supersymmetries doubled in the dimensional reduction process, as
the complex supercharges in four dimensions are converted into two
real supercharges in three dimensions.

Recall at this point that the $\CN=3$ and $\CN=4$ super Yang-Mills
equations are the same and only the field content differs by an
additional reality condition \cite{Witten:1978xx} in the case
$\CN=4$; this condition renders the fourth supersymmetry linear.
The $\CN=6$ and $\CN=8$ super Yang-Mills equations in three
dimensions are identical in the same sense.

There is a further Penrose-Ward transform for ordinary Yang-Mills
theory in four dimensions, which can also be translated to a
Yang-Mills-Higgs theory in three dimensions. Here, one considers
holomorphic vector bundles over a third-order thickening
of\footnote{All the spaces in the following are derived from the
corresponding superspaces by putting their fermionic coordinates
to zero.} $\CL^5\subset \CP^3\times \CP^3_*$. That is, instead of
demanding that $\kappa$ in \eqref{quadric} vanishes, we only
impose the condition that $\kappa^3\sim 0$ and arrive at an
infinitesimal neighborhood of $\CL^5$ in $\CP^3\times \CP^3_*$.
For a recent review on such complex manifolds which have
additional even nilpotent directions, see e.g.\
\cite{Saemann:2004tt}. The Penrose-Ward transform, which can then
be established \cite{Witten:1978xx,Isenberg:kk} maps holomorphic
vector bundles over the third-order thickening of
$\CL^5\subset\CP^3\times \CP^3$ to solutions to the ordinary
Yang-Mills equations on $\FC^4$.

The corresponding {\em mini-ambitwistor space} $\CL^4$ is obtained
by simply dropping all fermionic coordinates of $\CL^{4|6}$. For
the Penrose-Ward transform, one has in fact to consider a
``subthickening'', i.e.\ the space
$\CL^{4}\subset\CP^{2}\times\CP^{2}_*$ with the formal third-order
neighborhood of the diagonal $\Delta$ in the base $\CPP^1\times
\CPP^1_*$ of the fibration $\CL^4$ \cite{Saemann:2005ji}. We can
then establish a Penrose-Ward transform, which states that
holomorphic $\CL^4$-bundles over a third order subthickening of
$\CL^4$ which become holomorphically trivial vector bundles upon
restriction to any $\CPP^1\times \CPP^1_*\subset\CL^4$ are in a
one-to-one correspondence with solutions to the Yang-Mills-Higgs
equations on $\FC^3$ up to $\CL^4$-bundle equivalence and gauge
equivalence relations.

Summing up, we can state that there are twistor spaces for both
Yang-Mills-Higgs theory and $\CN=6$ super Yang-Mills theory in
three dimensions. Although these spaces are fibrations but no
manifolds, they come with nice properties, and the
mini-superambitwistor space $\CL^{4|6}$ possibly plays an
important r{\^o}le as the mirror manifold of the mini-supertwistor
space $\CP^{2|4}$. For future work, it remains to define a
topological B-model on the mini-superambitwistor space and to
substantiate the above pronounced mirror conjecture.

\end{document}